\documentclass[superscriptaddress,showpacs,twocolumn,prl,floatfix]{revtex4-1}
\usepackage{graphicx,graphics}
\usepackage{dcolumn}
\usepackage{amsmath,amssymb,amsfonts,latexsym,verbatim,mathrsfs}
\usepackage{bm,bbm}
\usepackage{color}

\def\be{\begin{equation}}
\def\ee{\end{equation}}
\def\ber{\begin{eqnarray}}
\def\eer{\end{eqnarray}}

\def\pv{{\bf p}}

\def\vv{{\bf v}}

\def\qv{{\bf q}}
\def\Av{{\bf A}}
\def\Ev{{\bf E}}
\def\Lv{{\bf L}}

\def\Pv{{\bf P}}

\def\ev{{\bf e}}

\def\vv{{\bf v}}

\def\Mv{{\bf M}}
\def\Dv{{\bf D}}
\def\Sv{{\bf S}}
\def\sigmabold{\mbox{\boldmath $\sigma$}}

\def\qv{{\bf q}}
\def\Av{{\bf A}}

\def\Ev{{\bf E}}
\def\Pv{{\bf P}}

\def\pv{{\bf p}}

\def\vv{{\bf v}}

\usepackage [english]{babel}
\usepackage{color}
\newcommand{\unit}[1]{\ensuremath{\, \mathrm{#1}}}

\begin{document}
\title{Electric Control of Spin Currents and Spin-Wave Logic}
\author{Tianyu Liu and G. Vignale}
\affiliation{Department of Physics and Astronomy, University of Missouri, Columbia, Missouri 65211, USA}
\begin{abstract}
{Spin waves in insulating magnets are ideal carriers for spin currents with low energy dissipation.   An electric field can modify the dispersion of spin waves, by directly affecting, via spin-orbit coupling, the electrons that mediate the interaction between magnetic ions.  Our microscopic calculations based on the super-exchange model indicate that this effect of the electric field is sufficiently large to be used to effectively control spin currents. We apply these findings to the design of a spin-wave interferometric device, which acts as a logic inverter and can be used as a building block for room-temperature, low-dissipation logic circuits.}
\end{abstract}
\maketitle

\begin{figure}
{\includegraphics[width=6.5cm]{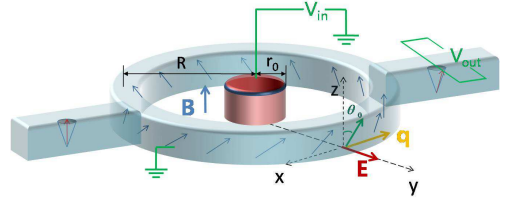}}
\caption{A Mach-Zehnder spin-wave interferometer in the presence of radial $\Ev$ field. A weak magnetic field is applied perpendicular to the ring plane, tilting the equilibrium magnetization away from the ring but still in the tangential plane to the ring. $\theta_0$ denotes the orientation of the equilibrium magnetization.} \label{FIGringwithE}
\end{figure}

One of the major challenges of contemporary electronics is to reduce dissipation as the size of devices shrinks to the nanometric scale.
In this context, spin-wave spintronics, so called magnonics, with insulating magnets offers interesting possibilities~\cite{Khitum, Kajiwara}. While in metals and semiconductors the spin current is carried by mobile
conduction electrons/holes, which inevitably dissipate energy as they move,
in a magnetic insulator, such as $Y_3Fe_5O_{12}$ (YIG), the spin current is carried by a collective motion of
magnetic moments -- a spin wave -- with no charge displaced.  The spin current
propagating in these insulating material is thus totally free of energy dissipation from
Joule heating, and almost free of dissipation from other sources (e.g. electron-magnon scattering):
 the coherence length can be as large as several centimeters~\cite{Kajiwara}.
For these reasons, magnetic insulators have attracted considerable attention in recent
theoretical~\cite{Schutz,Dugaev,Meier} and experimental~\cite{Onose,Kajiwara} work. For example, Kajiwara \textit{et al.}~\cite{Kajiwara}, have demonstrated injection and extraction of spin waves into and out of a YIG  wave guide~\cite{Serga}.
 Kostylev \textit{et al.}~\cite{Kostylev},  have designed an ingenious scheme of spin-wave logic, based on the interference between spin waves traveling along different arms of a Mach-Zehnder interferometer (a schematic illustration of a Mach-Zehnder spin wave interferometer is shown in Fig.~\ref{FIGringwithE}).

A crucial element of magnonics~\cite{Khitum} is the {\it phase shifter} -- a device that changes the phase of propagating spin waves. Several mechanisms have been proposed in the past to implement controlled phase shifts on spin waves. The simplest and most direct, is the application of a magnetic field, which shifts the dispersion~\cite{Landau}, thus changing the wave vector at constant frequency~\cite{Kostylev}. More sophisticated mechanisms exploited the Berry phase accumulated by spin waves that propagate on a non-collinear magnetic texture~\cite{Dugaev,Braun}. In a parallel development, Cao \textit{et al.}~\cite{Cao} studied the effect on spin waves of an electric field-induced Aharonov-Casher (AC) phase~\cite{Casher}. More recently, the influence of electric fields on spin waves has been studied both theoretically~\cite{Mills} and experimentally~\cite{Rovillain} and a  strong shift of spin-wave dispersion induced by an electric field has been reported~\cite{Rovillain}.

\begin{figure}
\includegraphics[width=3.8cm]{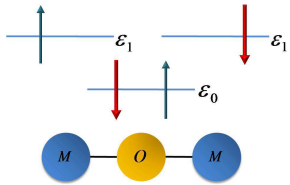}
\caption{Super-exchange model: two half-filled magnetic ions connected by an oxygen ligand. }
\label{FigSupex}
\end{figure}

 In this Letter we directly tackle the problem of controlling the phase of a spin wave (and hence the spin current) by means of an electric field. We will show that the electric field-induced AC phase has important implications for spin wave interferometry,  since the effect is much larger than that initially predicted in Ref.~\onlinecite{Cao}. Our analysis starts at the microscopic level, with a very simple super-exchange model~\cite{Mattis} for the magnetic interaction between two neighboring magnetic ions  (e.g. Fe$^{3+}$ in YIG) in a magnetic insulator. The model is depicted in Fig.~\ref{FigSupex}. There are no itinerant electrons in an insulator, but the virtual hopping of electrons between the  $d$-orbitals on the magnetic ions (Fe$^{3+}$) and the $p$-orbitals on the the ligand ($O^{2-}$) is sufficient to establish an antiferro-magnetic interaction of the Heisenberg type,
 \be\label{HH}
 H_H=J \Sv_1\cdot\Sv_2
 \ee
 which is responsible for the occurrence of magnetic order in the material.   Notice that in this model the physical $d$ and $p$ orbitals are replaced by doubly degenerate orbitals, which are eigenstates of the $z$-component of the electron spin.  Spin-orbit coupling effects are completely neglected up to this point.

 We now augment the usual super-exchange model  by the inclusion of a spin-orbit  (SO) interaction of the form
 \be\label{HSO}
 H_{SO}=-\frac{\lambda^2}{\hbar} (\pv \times e\Ev)\cdot\sigmabold\,,
 \ee
 where $\lambda$ is a characteristic length scale that controls the strength of the SO interaction, $e$ is the elementary charge and $\sigmabold$ is the Pauli matrix.  For electrons in vacuum $\lambda$ is the Compton wavelength $\lambda_c =\frac{\hbar}{mc}$, but we will see below that, in any realistic model of magnetic insulators, the value of $\lambda$ is orders of magnitudes larger (of the order of $\unit{\AA}$).  Although there are no itinerant electrons carrying a finite average momentum $\pv$, we will show below that the effect of the SO interaction on the phases of the virtual hopping translates, at the macroscopic level, into the appearance of a Dzyaloshinskii-Moriya (DM) interaction~\cite{Moriya} between the magnetic moments of the ions:
\be\label{HDM}
H_{DM}= \Dv\cdot (\Sv_1\times\Sv_2)
\ee
where the vector $\Dv$ is given by
\be\label{Dvector}
\Dv = -J \frac{ea}{E_{SO}} \Ev \times \hat \ev_{12}\,,
\ee
with $a$ being the distance between the magnetic ions, $\hat\ev_{12}$ the unit vector in the direction connecting the ions, and $E_{SO}\equiv \frac{\hbar^2}{2 m\lambda^2}$ (in vacuum $\lambda=\lambda_c$ and $E_{SO}=mc^2/2$, $m$ is the bare electron mass).
As a result of this interaction, the spin waves -- obtained by solving the appropriate {\it ferromagnetic} Hamiltonian for an infinite chain of identical blocks of magnetic ions (see discussion below) --  have their wave vector shifted by $\qv = \frac{\Dv \cdot {\bf \hat S_0}}{Ja}\hat\ev_{12}$, where ${\bf \hat S_0}$ is the direction of the equilibrium magnetization.  This, in principle, gives us a way to control the phase of the spin wave by an electric field.  In practice, the feasibility of the  proposal depends  critically on the strength of the spin-orbit coupling $\lambda^2$.  If we used the value of $\lambda$ in vacuum (as was done in Ref.~\onlinecite{Cao}), the effect would be extremely weak, and probably unobservable with realistic electric fields.

The reason why $\lambda$ turns out to be much larger than $\lambda_c$ is that the physical $d$-orbitals in the magnetic insulator (as opposed to the model orbitals we have dealt with so far) have strong intrinsic spin-orbit coupling $\Lv \cdot \Sv$ built in.  For example, in the model adopted by Katsura {\it et al.}~\cite{Katsura}, the doubly degenerate orbitals of the super-exchange model are actually spin-orbit-entangled states of the form $|a\rangle=(|d_{xy}\uparrow\rangle+|d_{yz}\downarrow\rangle+i|d_{zx}\downarrow\rangle)/\sqrt 3
$ and $|b\rangle=(|d_{xy}\downarrow\rangle-|d_{yz}\uparrow\rangle+i|d_{zx}\uparrow\rangle)/\sqrt 3
$.
In order to determine the value of $\lambda$ (or, better, $E_{SO}$) in our model, we observe that the DM interaction  Eqs.~(\ref{HDM},\ref{Dvector}), and indeed the spin-orbit interaction (\ref{HSO}) itself, can be viewed as the interaction of the electric field with an effective electric dipole, i.e. $H_{DM}=\Ev\cdot\Pv$, where the electric dipole is
\be\label{OurDipole}
\Pv = -J\frac{ea}{E_{SO}}\ev_{12} \times (\Sv_1 \times \Sv_2)\,.
\ee
The effective electric dipole arises from the hybridization of orbitals centered at different atoms. For example, in the model of Ref.~\cite{Katsura} the exact single-particle eigenstates are combinations of $d$ orbitals on the magnetic ions and $p$ orbitals on the oxygen due to the hopping of the electron: these states carry an electric dipole moment
$
\Pv = - \frac{4 e JI}{9t} \ev_{12}\times (\Sv_1 \times \Sv_2)\,,
$
where $t$ is the hopping coefficient and
$I=\frac{16}{27}Z_O^{5/2}Z_M^{7/2}(\frac{Z_O}{2}+\frac{Z_M}{3})^{-6}a_B$ with $a_B$ being the Bohr radius and $Z_O$ ($Z_M$) being the atomic number of {\bf O} ({\bf M}).
Comparing this to our Eq.~(\ref{OurDipole}) we arrive at an unambiguous identification of $E_{SO}$ (and hence $\lambda$) {\it within our model}:
\be\label{ESO}
E_{SO}=\frac{9ta}{4I}\,.
\ee
Taking YIG
 as an example~\cite{Lehmann,Cherepanov}, with $t=0.8$ eV, and $I=0.61a$, we get  $E_{SO}=3.0\unit{eV}$ and $\lambda=1.13\unit{\AA}$. This is indeed several orders of magnitude larger than $\lambda_c$  and opens the way to practical schemes of electric control of the phase of spin waves.

In the remaining part of this paper we supply more theoretical detail on the calculations supporting the above analysis, then work out the dispersion of spin waves in the presence of the electric field, and apply the results of these calculations to the design of a spin-wave interferometer (see Fig.\ref{FIGringwithE}), to be used as building block of spin-wave logic circuits.

{\em Microscopic analysis} -- The insulator we are interested in is YIG, whose magnetic order mainly arises from the super-exchange interaction
between $Fe^{3+}$ in octahedral (a) sites and tetrahedral (d) sites. The fact that the numbers of (a) and (d) sites per unit cell are different makes YIG a {\em ferrimagnet}. However, the long wave-length spin waves, whose energy is less than about $40$ K, can be understood with an effective {\it ferromagnetic}
exchange coupling between ``block spins" $S_i$, one per unit cell~\cite{Cherepanov}. The question is how an electric field affects this block ferromagnet. We start from the super-exchange model shown in Fig.~\ref{FigSupex}, which can be described by
the following Hamiltonian
\be\begin{aligned}
&H_{super}=H_0+H_t+H_U\,,\\
&H_0=\epsilon_0\sum_{\sigma}c_{0\sigma}^{\dag}c_{0\sigma}
    +\epsilon_1\sum_{i=1}^2\sum_{\sigma}c_{i\sigma}^{\dag}
    c_{i\sigma}\,,\\
&H_t=-t\sum_{\sigma}(c_{1\sigma}^{\dag}c_{0\sigma}+
    c_{2\sigma}^{\dag}c_{0\sigma}
    +h.c.)\,,\\
&H_U=U\sum_{i=1}^2\sum_{\sigma}c_{i\sigma}^{\dag}c_{i\sigma}c_{i\bar{\sigma}}^{\dag}c_{i\bar{\sigma}}\,,
\end{aligned}\ee
where $c$ ($c^{\dag}$) is the creation (annihilation) operator of
ligand electrons, which can hop forth and back only between oxygen
ligand and the metal ions, $\epsilon_1$ and $\epsilon_0$ are the
orbital energies of a metal ion and the oxygen ligand,
respectively.  The large repulsion energy $U$ ($\sim 8$ eV) between two electrons on the same metal ion  allows  for a maximum occupancy of two electrons per ion (the repulsion between the electrons in the oxygen ligand is negligible in comparison).

The fact that $t$ ($\simeq 0.8\unit{eV}$) is much smaller that $U$ allows us to use perturbation theory.
Keeping up to the fourth order of $t$ yields the effective interaction
between the spins on the magnetic ions:
\be
H_{eff}\simeq
\big(\frac{4t^4}{V^2U}+\frac{4t^4}{V^3}\big)\big[\frac{1}{2}(S_1^+S_2^-+h.c.)+S_1^zS_2^z\big]\,,
\ee
where $V=\epsilon_1-\epsilon_0+U$ corresponds to the energy difference between the
$p$ and  $|P\rangle_j$ orbitals in the paper by Katsura \emph{et al.}~\cite{Katsura}.  Setting  $J=\frac{4t^4}{V^2U}+\frac{4t^4}{V^3}\approx
\frac{8t^4}{V^3}$
and dropping the constant term, we obtain the Heisenberg interaction Eq.(\ref{HH}).
A positive $J$ implies that the interaction between neighboring magnetic ions is antiferromagnetic.   However, this antiferromagnetic interaction gives rise to a ferromagnetic interaction between ``block spins" in YIG, due to the unequal magnitudes of the anti-parallel magnetic moments in each block.

Let us now include the spin-orbit interaction $H_{SO}$ from Eq.~(\ref{HSO}).   It is easy to see that the inclusion of this interaction is equivalent to  the inclusion of a spin-dependent vector potential $\Av = \frac{m\lambda^2}{\hbar} \Ev \times \sigmabold$, which in turn modifies the hopping term $H_t$ by
a spin-dependent phase factor, that is
\be
H_t=-t\sum_{\sigma}(c_{1\sigma}^{\dag}c_{0\sigma}e^{-i\alpha\sigma}+
    c_{2\sigma}^{\dag}c_{0\sigma}e^{i\alpha\sigma}
    +h.c.)\,,
\ee
where $\alpha=\frac{eaE}{4E_{SO}}$, provided that the external electric field, the motion of the electron, and the electron spin ($\sigma$) are perpendicular to each other. Notice that the phase $\alpha$ is proportional to the distance between neighboring sites and independent of the direction of the localized moments: one can therefore switch to the ``block spins" description by simply reinterpreting $a$ as the distance between neighboring blocks.
 The resulting spin Hamiltonian takes the form
\ber\label{HSup}
H&=&-J'\sum_{<i,j>}S_i^zS_j^z+\frac{1}{2}(e^{i2\alpha_{ij}}S_i^+S_j^-+e^{-i2\alpha_{ij}}S_i^-S_j^+)\notag\\
&\simeq&-J'\sum_{<i,j>}\left\{( \Sv_i\cdot \Sv_j)+\sin2\alpha_{ij}(\Sv_i\times \Sv_j)_z\right\}\,,
\eer
where $\alpha_{ij} \equiv 2\alpha(i-j)$, $-J'$ is the effective exchange coupling for the spin blocks and $z$ is in the direction perpendicular to $\Ev$ and $\ev_{ij}$. In addition to the normal Heisenberg term we now also have a DM term~\cite{Moriya}, whose strength is linear in $E$.  An electric-field induced anisotropy is also present, but is  an effect of order $E^2$ and has therefore been neglected for weak electric field.

 In spite of the presence of the noncollinear DM term, the ferromagnetic configuration is still the ground state of~(\ref{HSup}).
 To show this, we make $\Sv_i= \Sv_0+\delta \Sv_i$, where $\delta \Sv_i$ is small deviation perpendicular to $\Sv_0$. Then the variation of the DM term up to the second order of
  $\delta \Sv_i$ is
 \be\label{DDM}
\delta H_{DM}\approx\sum_{<i,j>} \Dv_{ij}\cdot(\delta \Sv_i\times \Sv_0+\delta \Sv_i\times\delta \Sv_j)\,,
 \ee
 where $\Dv_{ij}$ is defined by Eq.(\ref{Dvector}) with $\ev_{12}$ replaced by $\ev_{ij}$. Since $\ev_{ij}= -\ev_{ji}$, we see that $\sum_{j}\Dv_{ij}=0$,
 which means $\delta H_{DM}=0$ up to the first order of $\delta \Sv_i$. Hence, the ground state is still ferromagnetic.  However, the DM term will definitely
 modify the spin-wave frequency, which involves a correction to the ground state energy at the second order in $\delta \Sv_i$. Further, we can clearly see  that it is only the component of the $\Dv$ parallel to $\delta \Sv_i\times\delta \Sv_j$ (i.e. to the direction of the equilibrium magnetization) that plays a role in the modification.

{\em  Spin wave dispersion} -- We now proceed to solve the dispersion of the spin waves in the presence of the DM interaction derived above. We consider the ring geometry illustrated in Fig.\ref{FIGringwithE}: the electric field perpendicular to the ring produces a DM vector $\Dv$ directed along the $z$-axis.  This will affect the dispersion of spin waves if and only if the equilibrium magnetization has a non-vanishing component along the $z$ axis.
In a flat ring, such as the one shown in Fig.\ref{FIGringwithE}, the shape anisotropy\cite{parameter} $-K(\Sv\cdot\hat\ev)^2/S^2$ where $\hat{\ev}$ is the unit vector along the ring -- outweighs other forms of anisotropy,  causing the equilibrium magnetization to lie along the ring, in which case the electric field has no influence. As a result, a magnetic field along the $z$ axis (Zeeman coupling $g\mu_B BS_z$) is necessary for us to observe the impact of the DM term on spin waves propagating in the ring.  Now, however, the orientation of the equilibrium magnetization is no longer constant in absolute space (even though it is constant relative to the ring).  This causes an additional geometric phase ($\alpha_g=\frac{a}{R}$) to appear, as shown in Ref.~\onlinecite{Dugaev}, where $R$ is the radius of the ring.  Putting everything together, i.e., DM interaction, geometric phase, Zeeman coupling and shape anisotropy, we arrive at the following equation of motion:
\ber
\hbar\frac{\partial\Sv_i}{\partial t}&=&\Sv_i\times[J'(\Sv_{i+1}+\Sv_{i-1})+2\frac{K}{S^2}S_i^x\hat{\ev}_x-g\mu_B B_i \hat{\ev}_z]\notag\\
&&-D_z S_i^z(\Sv_{i+1}-\Sv_{i-1})\,.
\eer
The large magnitude of the ``block spin" of YIG (S=14.3) allows us to use the semiclassical spin-wave approach to get the dispersion relation:
\be\label{dispersion}
\omega=\frac{J'S}{h}\big\{a^2\sqrt{(k^2+\kappa^2)(k^2+\kappa^2\sin^2\theta_0)}+ 2\bar{\alpha} ka\cos\theta_0\big\}\,,
\ee
where $\bar{\alpha}=4\alpha-\alpha_g$, $\kappa=\sqrt{\frac{K}{J'S^2a^2}}$, and $\cos\theta_0=\frac{g\mu_B BS}{2K}$, which
 is determined by minimizing the total Hamiltonian in the limit of $\kappa^2\gg\frac{1}{R^2}$.

\begin{figure}
{\includegraphics[width=7.2cm]{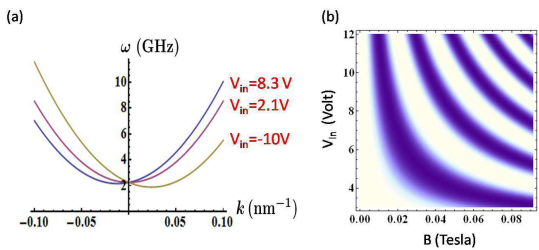}}
\caption{(a) Dispersion of spin waves in the ferromagnetic ring in Fig.\ref{FIGringwithE}, taking into account the geometric phase and the phase induced by the electric field. Parameters we use~\cite{parameter}: $J'=1.18\times10^{-4}\unit{eV}$, $K=1.53\times10^{-4}\unit{eV}$, $S=14.3$, $a=12.4\unit{\AA}$, $r_0=50\unit{nm}$, $R=100\unit{nm}$, $B=0.05\unit T$. (b) Transmission probability of a spin wave in the ring interferometer as a function of input voltage and magnetic field at $\omega=43\unit{GHz}$.} \label{FigDispersion}\label{FigTran}
\end{figure}

As shown in Fig.~\ref{FigDispersion}, one can tune the dispersion by adjusting the electric and the magnetic fields.  Just as a magnetic field shifts the spin wave dispersion {\it vertically} by increasing or decreasing the frequency at fixed $k$, the electric field shifts the dispersion {\it horizontally} by increasing or decreasing the wave vector at fixed frequency.

{\em Spin wave interferometer} -- Now we are ready to design our spin-wave interferometric device.  An insulating ring encircles a metal electrode to which a voltage $V_{in}$ can be applied.  The radial electric field acting upon the electrons in the ring is $\frac{-V_{in}}{R\ln(r_0/R)}$.

In Fig.~\ref{FigTran} (b)  we plot the transmission of a spin wave sent through this Mach-Zender interferometer, as a function of $V_{in}$ and B.  The effect of $B$ is to change the equilibrium orientation of the magnetization. The white regions in the figure are regions of constructive interference, separated by regions of destructive interference.
We see that very modest changes of potentials and magnetic fields, of the order of 1 V and 0.01 T respectively,  switch the response of the interferometer from high to low.  It is then clear how the device can be used as a logic inverter: the logic input being the voltage on the central electrode, and the logic output the intensity of the spin wave, as measured by an inductive coupler.
Advantages of this design are that it would operate at room temperature and GHz frequencies, with very little dissipation, and can be made small by using exchange spin waves -- the only type we are really considering here, since magnetostatic spin waves have much longer wavelengths and are hardly affected by the AC phase.
Once a logic inverter is available, we can follow Kostylev \emph{et al.}~\cite{Kostylev} in constructing more complicated architectures, which implement the NAND, the NOR, and all of classical logic.

It is worth noting that a traveling
spin wave is itself a source of electric field: $\Ev \simeq \mu_0\vv \times \delta \Mv$ where $\vv$ is the velocity of the spin wave, $\delta \Mv$ is the amplitude of the magnetization oscillation and $\mu_0=4 \pi\times 10^{-7}\unit{N/A^2}$ is the vacuum permeability. In our device, the resulting electric field is of the order of 1 V/m, which is negligible in comparison to the control field $\Ev \simeq 10^7$ V/m.  Even smaller is the electric field associated with mesoscopic equilibrium spin currents in the ring ($\Ev\simeq 10^{-2}$ V/m).~\cite{Schutz}

In conclusion, we have proposed an energy-efficient way to control the spin
current propagating in an insulating magnet by means of an electric field.
This possibility arises from the strong coupling that exists between the electric field and the spins of the electrons that mediate the interaction between magnetic ions.   The strength of this coupling has been theoretically estimated from microscopic
parameters, such as electron hopping coefficient, distance between neighboring magnetic sites, etc. as shown in Eq.(\ref{ESO}).  Or, it could be indirectly determined from measurements of physical effects that are sensitive to it, e.g. the spin wave spin Hall effect proposed by Meier and Loss~\cite{Meier}.  Finally, we have applied  our theory to an insulating magnetic ring inteferometer, which can be used to implement a voltage-controlled spin-wave-based NOT gate.

We acknowledge support from ARO Grant No. W911NF-08-1-0317 and thank Michael Flatt\'e for helpful discussions.


\begin{thebibliography}{99}
\bibitem{parameter}
    Shape anisotropy energy is $K=\mu_0M_s^2/2$, with $M_s=1.4\times 10^5\unit{A/m}$ being the saturation magnetization.
\end{thebibliography}
\end{document}